\documentclass[aps,showpacs, nofootinbib]{revtex4}

\newcommand{\be}{\begin{equation}}
\newcommand{\ee}{\end{equation}}
\newcommand{\ben}{\begin{eqnarray}}
\newcommand{\een}{\end{eqnarray}}

\newcommand{\la}{{\lambda}}

\newcommand{\si}{{\sigma}}

\newcommand{\cO}{{\cal O}}

\newcommand{\cM}{{\cal M}}

\newcommand{\cS}{{\cal S}}

\newcommand{\cL}{{\cal L}}

\newcommand{\tJ}{\tilde J}

\newcommand{\p}{\partial}
\newcommand{\na}{\nabla}

\newcommand{\tpsi}{\tilde \psi}
\newcommand{\tphi}{{\tilde \phi}}
\newcommand{\tchi}{\tilde \chi}

\newcommand{\tF}{\tilde F}
\newcommand{\tE}{\tilde E}
\newcommand{\tB}{\tilde B}
\newcommand{\tA}{\tilde A}
\newcommand{\tM}{{\tilde M}}

\newcommand{\tg}{\tilde g}
\newcommand{\tla}{\tilde {\lambda}}

\newcommand{\tR}{\tilde R}

\newcommand{\ep}{\epsilon}

\newcommand{\ga}{\gamma}

\newcommand{\talpha}{{\tilde \alpha}}
\newcommand{\tbeta}{{\tilde \beta}}


\begin{document}

\title{Angular momentum, mass and charge inequalities for black holes in Einstein-Maxwell gravity with dark matter sector}

\author{Marek Rogatko}
\affiliation{Institute of Physics \protect \\
Maria Curie-Sklodowska University \protect \\
20-031 Lublin, pl.~Marii Curie-Sklodowskiej 1, Poland \protect \\
marek.rogatko@poczta.umcs.lublin.pl \protect \\
rogat@kft.umcs.lublin.pl}

\date{\today}

\begin{abstract}
Angular momentum and mass-charge inequalities for axisymmetric maximal time-symmetric initial data in Einstein-Maxwell gravity
with {\it dark matter} sector were derived. The {\it dark matter} sector is mimicked by another $U(1)$-gauge field coupled to the ordinary Maxwell one.
We assume that data set
with two asymptotically flat regions is given on smooth simply connected manifold. 
One also pays attention to the area momentum charge inequalities for a closed orientable
two-dimensional spacelike surface embedded in the spacetime of the considered theory.
It turned out that the addition of {\it dark matter} sector influences to the great extent on the growth of the black hole masses. This fact may be responsible
for the existence of the recently detected supermassive objects at relatively short time after Big Bang. The inequalities binding the charge of the body (black hole) and its
size were also investigated. It turns out that the estimation depends on the coupling constant of the ordinary matter to the {\it dark matter} sector.
\end{abstract}

\maketitle

\section{Introduction}
Dynamical process of the formation of black holes is still a serious challenge to the present-day investigations both in general relativity and its extensions. Therefore,
from theoretical point of view it is very important to derive certain estimates and inequalities binding physical quantities characterizing process in question. 
Based on the first principles these formulae are of great importance for our understanding of these complicated processes.

First researches directed to this problem were given in \cite{pen73}, then generalized in the context of time-symmetric, axisymmetric data \cite{bri59, gib06} and developed
to obtain mass, angular momenta in the case of vacuum and electrovacuum \cite{dai06}-\cite{cos10}. On the other hand, quasi local quantities characterizing black holes
were considered in \cite{dai10}-\cite{dai11}, where the studies were conducted for both vacuum and cosmological constant spacetimes.
The extension of the above problems, including 
electric and magnetic charges into considerations was studied in \cite{jar11}-\cite{dai12}.
In \cite{dai13} the lower bound for the area of black holes in terms of their masses, charges and angular momenta was found, under the condition that the
initial data arouse from extreme Kerr-Newmann spacetime.

Another interesting aspects of the problem were studies of universal inequalities concerning size, angular momentum and charge of the body \cite{gab13}-\cite{ang16}.
Among these problems a general and sufficient conditions for a formation of black holes due to the concentration of angular momentum and charge were presented
\cite{khu15, khu15a}.

A natural extension of the researches is to answer the question concerning collapse and emergence of black objects in generalized theories of gravity as well as the higher
dimensional ones. The question concerning the influence of the other ingredients of our Universe like {\it dark matter} and dark energy 
constitutes an interesting research problem and should also be taken int account.
Some results concerning the uniqueness theorem of higher dimensional black objects \cite{gib02}-\cite{shi12}, existence of black holes in generalization of Einstein gravity \cite{mas93}-\cite{yaz16}, 
their masses and event horizon area \cite{hol12}-\cite{rog14a},  were achieved.

One of the key predictions of the current understanding of the creation of structures in the Universe is 
the so-called $\Lambda$CDM- model. It predicts that galaxies are embedded
in very extended and massive halos composed of {\it dark matter}, which in turns are surrounded 
by smaller {\it dark matter} sub-halos. The sub-halo {\it dark matter}
clumps are large enough to accumulate gas and dust and form satellite galaxies, which can orbit around the host ones. 
In principle smaller galaxies can be circled 
by much smaller sub-halo {\it dark matter} satellites, almost invisible by telescopes \cite{sta16}. So in the 
nearby of the Milky Way one can suppose that such structures 
also exist.
{\it Dark matter} interaction with the Standard Model particles
authorizes the main theoretical searches of the particle physics in the early 
Universe \cite{reg15}-\cite{foo15a}. It was also announce that  the collapse of neutron stars 
and emergence of the first star generations could contain some other hints for these researches \cite{bra14}-\cite{lop14}. 
The existence of {\it dark matter} can affect
black hole growth during the early stages of our Universe. The numerical studies of {\it dark matter} and dark energy 
collapse and their interactions with black holes and wormholes were studied in \cite{nak12,nak15a}.

On the other hand, physics processes such as 
increases of the gamma rays emissions coming from dwarf galaxies \cite{ger15},
possible dilaton-like coupling to photons caused by ultralight {\it dark matter} \cite{til15}, and
oscillations of the fine structure constant, can be explained by physics beyond the Standard Model.

There is still a great interest in observations of highly redshifted quasars, the potential place of location of supermassive black holes.
Approximately forty quasars at the distances greater than the redshift equal to six, with the mass of about one billion solar masses have been spotted. But this is only beginning.
Recent observations confirm the existence of black hole monsters of mass in the range between 12 to 17 billion solar masses \cite{fan13}-\cite{tho16}. 
The existence of such supermassive black objects in the Early Universe when its age was less than one billion years is a big challenge for the theory of black hole formation, growth and coevolution with
early galaxies. The tantalizing question is how in a relatively short time after Big Bang such huge concentration of masses can be explained.

Motivated by the above statements concerning the {\it dark matter} and its possible influence on the Standard Model physics, we shall look for the lower bound for area 
and mass of the black holes in Einstein-Maxwell {\it dark matter} gravity. In our considerations we shall elaborate the model of {\it dark matter} sector, which
the ordinary Maxwell field is coupled to the other $U(1)$-gauge field mimicking {\it dark matter}. The model in question has strong astronomical and Earth experiment motivations like
observations of $511$ eV gamma rays, electron positron excess in galaxy, muon anomalous magnetic moment explanation, dark photon detection experiments \cite{integral} -\cite{babar14}. 

The organization of our paper is as follows. In Sec.2 we describe the basic features of Einstein-Maxwell gravity with {\it dark matter} sector. Section 3 will be devoted to the initial data for the underlying theory.
In Sec.4 we pay attention to the properties of the angular momentum, while Sec.5 is devoted to the mass inequalities for black holes with {\it dark matter}. It turns out that the participation of this ingredient of the 
Universe mass in gravitational collapse can be treated as an efficient factor for explanation of the huge masses of primordial black objects.
Section 6 is concerned with the area inequalities, while in Sec.7 we have given some remarks concerning the charge and the size of the body with the influence of the {\it dark matter} sector.
Firstly one assumes the maximal initial data and in the next step we try to get rid of the aforementioned assumption.
In Sec.8 we concluded our investigations.

\section{Gravity with dark matter sector}
In this section we shall elaborate the basic features of Einstein-Maxwell gravity supplemented by the {\it dark matter}  sector, which constitutes another $U(1)$-gauge field coupled
to the ordinary Maxwell one. 
Recently, the justification of the model in question was given from string/M-theory \cite{ach16}.  The mixing term of the two gauge fields, called {\it kinematic mixing portals} is typical
for states of open string theories, i.e., both gauge states are supported by D-branes. They are in turn separated in the extra dimensions like in supersymmeric Type I, Type IIA and Type IIB models.

The elaborated model with {\it dark matter } sector was also widely studied in the AdS/CMT attitude, as a possible influence of the {\it dark matter} on condensed matter systems, being a potential
guidelines for future experiments \cite{adscmt}.  As far as the latest {\it experimental} search for the dark photons is concerned, constraints  on their masses below $\sim 100$ MeV from the
observations of Supernova 1987A were given in Ref.\cite{cha17}.

The action of the theory into consideration is provided by 
\be
S = \int \sqrt{-g}~ d^4x  \bigg( \frac{R}{16 \pi}
-  \frac{1}{4}F_{\mu \nu} F^{\mu \nu} - \frac{1}{4} B_{\mu \nu} B^{\mu \nu} - \frac{\alpha}{4} F_{\mu \nu} B^{\mu \nu}
\bigg), 
\ee
where
$F_{\mu \nu} = 2 \nabla_{[ \mu} A_{\nu ]}$ stands for the ordinary Maxwell field strength tensor, while
the second $U(1)$-gauge field $B_{\mu \nu}$ is given by $B_{\mu \nu} = 2 \nabla_{[ \mu} B_{\nu ]}$,
$\alpha$ is a coupling constant between $U(1)$ fields. 

Varying the action with respect 
to the considered fields one obtains the equations of motion which can be written in the form as
\ben \label{eq1}
G_{\mu \nu} &=& 8 \pi~(T_{\mu \nu}(F) + T_{\mu \nu}(B) 
+ \alpha~T_{\mu \nu}(F,~B)) + 8 \pi~T_{matter},\\
\na_{\mu}F^{\mu \nu} &+& \frac{\alpha}{2}~\na_\mu B^{\mu \nu} = 0,\\ \label{eq2}
\na_{\mu}B^{\mu \nu} &+& \frac{\alpha}{2}~\na_\mu F^{\mu \nu} = 0.
\een
The contributions to the full energy momentum tensors are given by 
\ben
T_{\mu \nu}(F) &=& \frac{1}{2}~F_{\mu \beta}F_{\nu}{}^{\beta} - \frac{1}{8}~g_{\mu \nu}~F_{\alpha \beta}F^{\alpha \beta},\\
T_{\mu \nu}(B) &=& \frac{1}{2}~B_{\mu \beta}B_{\nu}{}^{\beta} - \frac{1}{8}~g_{\mu \nu}~B_{\alpha \beta}B^{\alpha \beta},\\
T_{\mu \nu}(F,~B) &=& \frac{1}{2}~F_{\mu \beta}B_{\nu}{}^{\beta} - \frac{1}{8}~g_{\mu \nu}~F_{\alpha \beta}B^{\alpha \beta}.
\een

\section{Initial data for Einstein-Maxwell dark matter theory}

To commence with we comment on the initial data for Einstein-Maxwell gravity coupled to {\it dark matter} sector.  In what follows one
supposes that we take into account non-electromagnetic matter fields. 

The globally hyperbolic manifold of the considered theory we foliate
by Cauchy surfaces $\Sigma_t$, parameterized by a global time $t$. Let  us suppose that $n_\alpha$
is the unit normal to the hypersurface in question. Then, one gets that $n_{\alpha}~n^\alpha = - 1$.
The spacetime metric induced on $\Sigma_t$ hypersurface, $h_{\alpha \beta}$, is provided by the relation
\be
h_{\alpha \beta} = g_{\alpha \beta} + n_\alpha n_\beta.
\ee
Moreover, 
we define {\it electric} and {\it magnetic} components for gauge field strengths $F_{\alpha \beta}$ and respectively for
$B_{\alpha \beta}$.  The {\it electric} components for the Maxwell and {\it dark matter} gauge field components imply
\be
E^{\alpha} =  F^{\alpha \beta}~n_\beta, \qquad
\tE^{\alpha } = B_{\alpha \beta}~n_\beta,
\ee
while the {\it magnetic} ones can be written in the forms provided by
\be
B^{\alpha } = - \ast F^{\alpha \ga}~n_\ga, \qquad
\tB^{\alpha } = - \ast B^{\alpha \ga}~n_\ga.
\ee
where one denotes the dual Maxwell tensor and {\it dark matter} one, respectively by
\be
\ast F_{\alpha \beta} = \frac{1}{2}~\ep_{\alpha \beta \ga \delta}~F^{\ga \delta}, \qquad
\ast B^{\alpha \beta} = \frac{1}{2}~\ep_{\alpha \beta \ga \delta}~B^{\ga \delta}.
\ee
The constraint equations 
which define time-symmetric initial data for the theory in question, are given by
\ben \label{c1}
D^a \bigg( K_{ab} - K_c{}^c~h_{ab} \bigg) &=& 4 \pi~\ep_{mib} \bigg[ B^m E^i + \tB^m \tE^i + \\ \nonumber
&+& \frac{\alpha}{2}~(B^m \tE^i + \tB^m E^i) \bigg] + 8 \pi P_b,\\ \label{c2}
{}^{(3)} R + (K_i{}^i)^2 - K_{mn}~K^{mn}  &=& 4 \pi~\bigg[
B_m B^m + E_a  E^a + \tB_m \tB^m +  \tE_a \tE^a\\ \nonumber
 &+& \alpha~(B_m \tB^m + E_a \tE^a) \bigg] + 16 \pi \mu_b,
\een
where $\mu_b$ is the non-electromagnetic matter energy density and $P_b$ denotes non-electromagnetic matter momentum density. $D_a$ is the derivative with respect to the metric tensor $h_{ab}$.

In the consideration in question we shall tackle the problem of asymptotically flat Riemannian manifold with a region diffeomorphic to
 $R^3 \setminus B(R)$, where $ B(R)$ is a coordinate ball of radius $R$. In 
 the aforementioned region we suppose that the following conditions will be quarantined:
\ben 
h_{ij} &-& \delta_{ij} = \cO_k(r^{-{1 \over 2}}), \qquad
\p_k h_{ij} \in L^2(M_{ext}), \qquad K_{ij} = \cO_{l-1}(r^{-3}), \\ \nonumber
E^i &=& \cO_{l-1}(r^{-2}), \qquad
\tE^i = \cO_{l-1}(r^{-2}), \\
B^i &=& \cO_{l-1}(r^{-2}), \qquad
\tB^i = \cO_{l-1}(r^{-2}),
\een
where we set $f = \cO_k(r^\la), ~\p_{k_1 \dots k_l} f = \cO(r^{\la -l})$, for
$0 \leq l \leq k$. 

The initial data set for Einstein-Maxwell gravity with {\it dark matter} sector will consist of metric tensor, extrinsic curvature, as well as 
vectors bounded with both gauge fields $(M,~h_{ij},~K_{ab},~E_i,~\tE_i,~B_a,~\tB_a)$.
We shall take into account the axisymmetric initial data, which means that the data will be invariant under the action of $U(1)$-group. 
Namely, the defined quantities should be invariant under the action of the group. In view of these conditions one has that
\be
\cL_\eta h_{ab} = \cL_\eta K_{ij} = \cL_\eta E_i = \cL_\eta \tE_i = \cL_\eta B_j
= \cL_\eta \tB_j = 0,
\label{sym}
\ee 
where $\cL_\eta$ is Lie derivative with respect to the Killing vector field $\eta_\alpha$.

\section{Angular momentum}
In this section the analysis will be addressed to the problem of an angular momentum in Einstein-Maxwell {\it dark matter} gravity theory. The angular momentum directed by 
the rotation axis of two-dimensional surface $\Sigma \in M$, with
a tangent vector $\eta_a$ and unit outer normal vector over the coordinate sphere, may be cast in the form
\be
J(\Sigma) = \frac{1}{8\pi}~\int_{\Sigma} d\Sigma ~(K_{ab} - K_c{}^c h_{ab})~n^a~\eta^b .
\ee
The above relation describes the so-called Komar-like angular momentum of the two-dimensional surface.
This definition coincides with the Komar one in the case when the tangent vector $\eta_a$ is defined in the adjacent of the surface $\Sigma$.
Unfortunately, the defined quantity is not necessary conserved. The problem is that some matter fields may have the bulk contribution defined in the form of the
Stoke's theorem and they can be bounded with black hole (surface terms). 

All the quoted arguments lead to the conclusion that one needs to modify the definition of the angular momentum, in order to remedy the situation in question.
The form of the total angular momentum is motivated by the demand to have a property of being conserved. Of course, one should take into account the contribution of Maxwell 
and {\it dark matter} sector. In the theory under inspection, 
the total angular momentum of a surface $\Sigma$ is provided by
\ben
\tJ(\Sigma) &=& \frac{1}{8\pi}~\int_{\Sigma} d\Sigma ~(K_{ab} - K_c{}^c h_{ab})~n^a~\eta^b + \frac{1}{2}\int_{\Sigma} d\Sigma ~A^b \eta_b~E^k  n_k\\ \nonumber
&+& \frac{1}{2}\int_{\Sigma} d\Sigma ~\tA^b \eta_b~\tE^k n_k  + \frac{\alpha}{4}\int_{\Sigma} d\Sigma ~A^b \eta_b~\tE^k n_k \\ \nonumber
&+& \frac{\alpha}{4}\int_{\Sigma} d\Sigma ~\tA^b \eta_b~ \tE^k n_k.
\een
It should be remarked that the topology of the manifold of the considered theory does not allow for globally smooth vector potentials.
In order to overcome this problem one ought to remove the Dirac string bounded with each point  $i_n$. Removing from the manifold in question, either the portion
of the $z$-axis below or above $i_n$, in order to have $U(1)$-invariant potentials. Namely, we have
\be
A_i = \frac{1}{ 2 M}~\sum\limits_{k=1}^M
\bigg( A^{(k)}_{+ i} + A^{(k)}_{- i} \bigg), \qquad B_i = \frac{1 }{ 2 N}~\sum\limits_{k=1}^N
\bigg( B^{(k)}_{+ i} + B^{(k)}_{- i} \bigg),
\ee
on $R^3  \setminus \{ z-axis \}$.  As was mentioned in \cite{dai13}, $A_i$ and $B_i$ are discontinuous on $z$-axis, but the product
$A_i~\eta^i$ and $B_a~\eta^a$ is well behaved. It happens due to the fact that the Killing vector  field $\eta_b$ vanishes on $z$-axis.

Let us quote some implications concerning the total angular momentum. To commence with, we consider the left-hand side of the constraint equation (\ref{c1}), extract $P_b$
and calculate an integral over $M_1$ from $P_b \eta^b$, where $M_1 \in M$ and its boundaries are given by $\p M_1 = \Sigma_1 \cup \Sigma_2$.
The terms can be further cast in the forms which yield
\ben
\ep_{bmi}~B^m  E^i \eta^b &=& D_k \bigg[ \ep_{bmi} E^i ~\ep^{mkl} A_i \eta^b \bigg] - D_i E^i ~\eta^b A_b,\\
\ep_{bmi}~\tB^m  E^i \eta^b &=& D_k \bigg[ \ep_{bmi} \tE^i ~\ep^{mkl} \tA_i \eta^b \bigg] - D_i \tE^i ~\eta^b \tA_b,\\
\ep_{bmi}~B^m  \tE^i \eta^b &=& D_k \bigg[ \ep_{bmi} \tE^i ~\ep^{mkl} A_i \eta^b \bigg] - D_i \tE^i ~\eta^b A_b,\\
\ep_{bmi}~\tB^m  E^i \eta^b &=& D_k \bigg[ \ep_{bmi} E^i ~\ep^{mkl} \tA_i \eta^b \bigg] - D_i E^i ~\eta^b \tA_b,
\een
where we here used the fact that $\cL_\eta E^a =0, ~\cL_\eta \tE^a = 0$. Having in mind equations of motion
\ben
D_i E^i &+& \frac{\alpha}{2}~D_i \tE^i = 0,\\
D_m \tE^m &+& \frac{\alpha}{2}~D_m E^m = 0,
\een
after straightforward calculations we arrive at the relation
\ben
\int_{M_1} dV~P^b \eta_b &=& \frac{1}{8\pi}~\int_{\p M_1} d\Sigma ~(K_{ab} - K_c{}^c h_{ab})~n^a~\eta^b + \frac{1}{2}\int_{\p M_1} d\Sigma ~A^b \eta_b~E^k n_k \\ \nonumber
&+& \frac{1}{2}\int_{\p M_1} d\Sigma ~\tA^b \eta_b~\tE^k  n_k + \frac{\alpha}{4}\int_{\p M_1} d\Sigma ~A^b \eta_b~\tE^k n_k \\ \nonumber
&+& \frac{\alpha}{4}\int_{\p M_1} d\Sigma ~\tA^b \eta_b~\tE^k  n_k\\ \nonumber
&=& \tJ(\Sigma_2) - \tJ(\Sigma_1).
\een
If $P_b \eta^b$ is equal to zero, we get the conclusion that the total angular momentum in Einstein-Maxwell {\it dark matter} gravity is conserved.

Let us ask the question about the gauge invariance of the total angular momentum, namely we shall check if $\tJ(\Sigma)$ is invariant under the the gauge transformation of the
$U(1)$ fields appearing in the theory. In order to show this we calculate the expression
\be
\tJ(\Sigma) = \tJ(\Sigma, ~A_i + \na_i a,~B_i + \na_i b),
\ee
and use the fact that $\eta_a n^a = 0$, as well as, $\cL_\eta (E^K~n_k) = 0,~\cL_\eta (\tE^K~n_k) = 0$,
the gauge invariance of the total angular momentum can be established.

As far as the gauge fields in theory is concerned, we have the following asymptotic behaviors:
\be
A_\mu \sim \cO \bigg(\frac{1}{r} \bigg), \qquad B_\mu \sim \cO \bigg(\frac{1}{r} \bigg),
\ee
and for the $U(1)$ Killing vector field $\eta_a$, one has
\be
\mid \eta \mid \sim \mid x~\p_y - y~\p_x \mid = \cO(\rho),
\ee
where $\rho \sim \sqrt{x^2 + y^2}$. In the case of {\it magnetic} and {\it electric} part of Maxwell and {\it dark matter} sector, we suppose
that their asymptotic expansions are given by
\be
(E_a,~\tE_a,~B_a, ~\tB_a) \sim \frac{1}{r^2}~\p_r + \cO \bigg( \frac{1}{r^3} \bigg).
\ee
On the other hand, it can be shown \cite{dai13} that the following relations are satisfied:
\be
\lim_{r \rightarrow \infty}
\frac{1}{  r^2}~
\frac{1}{ 2 M}~\sum\limits_{k=1}^M 
 \bigg( A^{(k)}_{+ i} + A^{(k)}_{- i} \bigg)\eta^i = 0, \qquad
\lim_{r \rightarrow \infty}
\frac{1}{  r^2}~
\frac{1}{  2 N}~\sum\limits_{k=1}^N
 \bigg( B^{(k)}_{+ i} + B^{(k)}_{- i} \bigg)\eta^i = 0.
\ee
It leads to the conclusion that the total angular momentum $\tJ$ tends at infinity
as $J(\Sigma)$ - Komar like angular momentum. The conclusions bounded with our calculations of the angular momentum in the theory in question can be formulated as follows:

\noindent
{\bf Theorem}\\
Let us suppose that $(M, ~h_{ij},~K_{ij},~E_i,~\tE_i,~B_i,~\tB_i)$ is the initial axisymmetric data for Einstein-Maxwell {\it dark matter} theory. Moreover, if one 
has that $P_m \eta^m = 0$ then the total angular momentum $\tJ(\Sigma)$ is conserved. It means that for $U(1)$ invariant
hypersurfaces $\Sigma_1$ and $\Sigma_2$ the following relation is satisfied:
\be
\tJ(\Sigma_1) = \tJ(\Sigma_2).
\ee
The total angular momentum is gauge invariant for both gauge fields (Maxwell and {\it dark matter}) and vanishes in a neighborhood of $i_n$ points, as well as, for $S_\infty$, one has that 
\be
\tJ(S_\infty) = J(\Sigma).
\ee

In what follows we shall consider the problem of the so-called {\it twist potential}. Having in mind the considerations connected with this subject in
ordinary Einstein-Maxwell gravity, let us define the following quantity:
\be
\la = \ep_{abc}~\bigg( \pi^{bk} - 4\pi~\Theta^{bk} \bigg)~\eta^c ~\eta_k~dx^a,
\label{twist}
\ee
where we have denoted by
\be
\pi_{ab} = K_{ab} - K_{c}{}{}^{c}~h_{ab}, 
\ee
and  for $\Theta_{ab}$ stands the following expression:
\be
\Theta_{ab} = \psi_{ab}^{(E)} + \psi_{ab}^{(\tE)} + \psi_{ab}^{(B \tA)} + \psi_{ab}^{(\tB A)}.
\label{pot}
\ee
On the other hand, the adequate terms in the relation (\ref{pot}) are provided by
\ben
\psi_{ab}^{(E)}  &=& \ep_{imb} E^i ~\ep_{a}{}^{lm} A_l, \qquad \psi_{ab}^{(\tE)}  = \ep_{imb} \tE^i ~\ep_{a}{}^{lm} \tA_l, \\ \nonumber
\psi_{ab}^{(B \tA)}  &=& \ep_{imb} E^i ~\ep_{a}{}^{lm} \tA_l, \qquad  \psi_{ab}^{(\tB A)}  = \ep_{imb} \tE^i ~\ep_{a}{}^{lm} A_l.
\een
In order to find the properties of the {\it twist potential} let us calculate $(d \la)_{ij}$.  Namely, one arrives at
 \be
 (d \la)_{ij} = D^a \bigg( \pi_{ab} ~\eta^b - 4 \pi~ \Theta_{ab} ~\eta^b \bigg) ~\ep_{ijl}~\eta^l.
 \ee
 Using the equations of motion for the underlying theory
 \ben
 D_i E^i &+& \frac{\alpha}{2}~D_i \tE^i = 0,\\
 D_i \tE^i &+& \frac{\alpha}{2}~D_i E^i = 0,
 \een
 and collecting various terms in order to obtain the adequate Lie derivatives multiplied by component of the gauge field potentials
 \be
 \cL_{\eta} E^m~A_m =0, \qquad \cL_{\eta} \tE^j ~\tA_j = 0,
 \ee
 we arrive at the expression $8\pi~P_b \eta^b~\ep_{ijk} \eta^k$. Just, for the condition of $P_b \eta^b = 0$, we  draw a conclusion that
 the {\it twist potential} form is closed, i.e.,
 \be
 (d \la)_{ij} = 0.
 \ee
 At the beginning of the considerations we also assume that the manifold in question is simply connected. It leads to the statement that 
 {\it twist potential}, as described by equation (\ref{twist}), exists.

\section{Mass inequalities for black holes with dark matter}
This section will be devoted to the investigation of the inequalities binding masses of the black objects in the considered theory.
In order to tackle the problem of black hole mass in Einstein-Maxwell {\it dark matter} gravity let us consider an axisymmetric line element which yields \cite{gib06,chr08a,chr08b}
\be
ds^2 = q_{AB}~dx^A dx^B + X^2~\bigg( d\varphi + W_B~dx^B \bigg)^2,
\ee
where the metric tensor $q_{AB}$ describes a two-dimensional orbit space of Killing vector
$\eta_\alpha = ( \p/\p \varphi)_\alpha$, while functions $X$ and $W_B$ are independent on
$\varphi$-coordinate.  On the other hand, when one requires the {\it strongly axisymmetric} condition to be satisfied, the additional mirror symmetry 
should be implemented causing that $W_B$ disappears.

The required properties of the metric are satisfied by
\be
ds^2 = e^{-2U + 2\beta}~(d\rho^2 + dz^2) + \rho^2~e^{- 2U} ~\bigg(
d \varphi  + \rho~W_\rho~d\rho + W_z~dz\bigg)^2.
\label{metr}
\ee
All the functions are $\varphi$-independent and the above choice of the coordinates corresponds to obtaining a harmonic function on the orbit space, i.e.,
Laplace operator with respect to the metric $q_{AB}$ acting on $\rho$ is equal to zero. 
The solution of the aforementioned equation is unique and one specifies conditions at infinity and $z$-axis, which
constitutes the boundary of the orbit space. Furthermore, the regularity of the axisymmetric metric imposes the additional requirements for function $U$ and $\beta$.

In the above defined coordinates we shall find the ADM-mass which can be written as
\cite{gib06,chr08a,chr08b} 
\be
m = \frac{1}{16 \pi} \int dx^3
~\bigg[
{}^{(3)}R + {1 \over 2} ~\rho^2 ~e^{- 4 \beta + 2 U}~\bigg(
\rho~W_{\rho,z} - W_{z, \rho} \bigg)^2 \bigg]~e^{2 \beta - 2 U}
+ \frac{1}{ 8 \pi}~\int dx^3~(D U)^2.
\label{mas}
\ee
The initial data invariant under the flow of the Killing vector $\eta_a = (\p /\p \phi)_a$ yield that the following relations are fulfilled:
\ben 
\cL_\eta F_{ab} &=& 0, \qquad \cL_\eta \ast F_{ab} = 0,\\
\cL_\eta B_{ab} &=& 0, \qquad  \cL_\eta \ast B_{ab} = 0,
\een
where the star stands for the Hodge'a dual.\\
Moreover the fact that the manifold under consideration is simply connected, ensures the existence of the potentials $\phi,~\tphi,~\psi,~\tpsi$, bounded with the adequate
$U(1)$-gauge field
\ben \label{pot1}
\na_\mu \phi &=& F_{\alpha \mu} ~\eta^{\alpha}, \qquad \na_\mu \psi = \ast F_{\mu \alpha } ~\eta^{\alpha},\\ \label{pot2}
\na_\mu \tphi &=& B_{\alpha \mu}~\eta^{\alpha}, \qquad \na_\mu \tpsi = \ast B_{\mu \alpha } ~\eta^{\alpha}.
\een
In the orthonormal basis of the line element (\ref{metr}) the adequate left-hand sides of the relations (\ref{pot1})-(\ref{pot2}), are provided by
\be
\p_\alpha \Phi = \sqrt{g_{\varphi \varphi}}~\Theta_{3 \alpha},
\ee
where $\Phi = (\phi,~\tphi,~\psi, ~\tpsi)$
and $\Theta_{3 \alpha} = (F_{\alpha \mu},~\ast F_{\alpha \mu},~B_{\alpha \mu},~\ast B_{\alpha \mu})$.

In \cite{dai08} it was found that the twist potential was bounded with the extrinsic curvature tensor 
$K_{ij}$ by the relation of the form
\be
\omega = 2~\ep_{ijk}~K^j{}{}_l~\eta^k~\eta^l~dx^i,
\ee
which enables to find that 
\be
e^{2 \beta - 2U} ~\mid K \mid_{h}^2 ~
\ge 2~e^{2 \beta - 2U}~(K_{13}^2 + K_{23}^2) 
= {e^{4U} \over 2~\rho^4}~\mid \omega \mid_{h}^2.
\label{costa}
\ee
We also assume that the initial data set in maximal, i.e., $K_{j}{}{}^j = 0$.
Then, we insert equation (\ref{costa}) into relation (\ref{mas}). The outcome is provided by
\ben \label{inemass}
m &\ge& {1 \over 16 \pi} \int dx^3~
\bigg[ {}^{(3)}R~e^{2 \beta - 2U} + 2~(DU)^2 \bigg] \\ \nonumber
&\ge& {1 \over 16\pi} \int dx^3 ~\bigg[
(DU)^2 + {e^{4U} \over 2~\rho^4}~\mid \omega \mid_{h}^2 + 
4 \pi ~e^{2(\beta - U)} ~\bigg(
B_m B^m + E_a  E^a + \\ \nonumber
&+& \tB^m \tB^m +  \tE_a \tE^a
 +\alpha~(B^m \tB^m + E_a \tE^a) \bigg) \bigg].
 \een
The third term in the bracket can be reduced to the form which yields
\ben
\frac{1}{g_{\phi \phi}} \bigg[
(\p_n \phi)^2 + \mid D \phi \mid_h^2  &+& 
(\p_n \psi)^2 + \mid D \psi \mid_h^2 + \\ \nonumber
+ (\p_n \tphi)^2 + \mid D \tphi \mid_h^2 &+& 
(\p_n \tpsi)^2 + \mid D \tpsi \mid_h^2  + \\ \nonumber
+ \alpha ~\bigg( (\p_n \psi)(\p_n \tpsi)+ \mid D \psi \mid_h ~\mid D \tpsi \mid_h &+& 
(\p_n \phi)(\p_n \tphi)+ \mid D \phi \mid_h ~\mid D \tphi \mid_h \bigg) \bigg] \\ \nonumber
\geq \frac{e^{2 U}}{\rho^2}~
\bigg[ \mid D \phi \mid_h^2 + \mid D \psi \mid_h^2 &+&  \mid D \tphi \mid_h^2 + \mid D \tpsi \mid_h^2  \\ \nonumber
+ ~\alpha ~\bigg( 
\mid D \psi \mid_h ~\mid D \tpsi \mid_h &+& \mid D \phi \mid_h ~\mid D \tphi \mid_h \bigg) \bigg],
\een
where $\p_n$ stands for the derivative taken in the direction of the unit normal to the initial data hypersurface.

Inspection of the equations of motion and the exact form of the energy momentum tensors enable us to rewrite the twist for the stationary Killing field $\eta^a$ in the form as
\be
\omega_{[a;b]} =  {4} \bigg[
\na_{[a} \psi \na_{b]} \phi + \na_{[a} \tpsi \na_{b]} \tphi + \frac{\alpha}{2}~\bigg( \na_{[a} \tpsi \na_{b]} \phi + \na_{[a} \psi \na_{b]} \tphi \bigg) \bigg].
\ee
It enables us to write the following relation:
\ben \label{diff}
d \bigg[ \frac{1}{2}\omega &-& (\psi~ d\phi - \phi~d \psi) - (\tpsi~ d\tphi - \tphi~d \tpsi) 
- \frac{\alpha}{2}(\tpsi~ d\phi - \phi~d \tpsi)  \\ \nonumber
&-& \frac{\alpha}{2}(\psi~ d\tphi - \tphi~d \psi) \bigg] = 0.
\een
The fact that the differential form described by the equation (\ref{diff}) is closed allows us to conclude that there is a function $v$ (determined up to a constant),
such that
\ben
\frac{1}{2} \omega = dv &+& (\psi~ d\phi - \phi~d \psi) + (\tpsi~ d\tphi - \tphi~d \tpsi) 
+ \frac{\alpha}{2}(\tpsi~ d\phi - \phi~d \tpsi)  \\ \nonumber
&+& \frac{\alpha}{2}(\psi~ d\tphi - \tphi~d \psi).
\een
Consequently the equation (\ref{inemass}) can be rearranged in the form as follows:
\ben
\label{inemass1}
m &\ge& {1 \over 8 \pi} \int dx^3~\bigg[
(DU)^2  + {e^{4U} \over \rho^4}~\mid 
dv + (\psi d\phi - \phi d \psi) + (\tpsi d\tphi - \tphi d \tpsi)  + \\ \nonumber
&+& \frac{\alpha}{2} (\tpsi d\phi - \phi d \tpsi) + \frac{\alpha}{2} (\psi d\tphi - \tphi d \tpsi)
\mid_{h}^2 + \\ \nonumber
&+& 2 \pi ~\frac{e^{2 U} }{\rho^2}~
\bigg( \mid D \phi \mid_h^2 + \mid D \psi \mid_h^2 +  \mid D \tphi \mid_h^2 + \mid D \tpsi \mid_h^2  \\ \nonumber
&+& ~\alpha ~( 
\mid D \psi \mid_h ~\mid D \tpsi \mid_h + \mid D \phi \mid_h ~\mid D \tphi \mid_h ) \bigg) \bigg],
\een
Further, let us define the total charges connected with the considered gauge fields. Namely, one has
\ben
Q_E &=& - \frac{1}{4 \pi} \int~\ast F_{\alpha \beta}~dS^{\alpha \beta}, \qquad Q_{\tE} = - \frac{1}{4 \pi}\int~\ast B_{\alpha \beta}~dS^{\alpha \beta}, \\
Q_M &=& \frac{1}{4 \pi}\int~F_{\alpha \beta}~dS^{\alpha \beta}, \qquad Q_{\tM} = \frac{1}{4 \pi} \int~B_{\alpha \beta}~dS^{\alpha \beta},
\een
The above definitions and the procedure presented in \cite{chr08a}-\cite{cos10} enables us to arrive at the inequalities connecting ADM mass and angular momentum of black hole
and the charges appearing in the Einstein-Maxwell {\it dark matter} gravity. In the next step, let us define the action of the form as
\ben
I = \int dx^3 \bigg[
 (DU)^2  &+& {e^{4U} \over \rho^4} \mid 
 dv + (\psi d\phi - \phi d \psi) + (\tpsi d\tphi - \tphi d \tpsi)  + \\
&+& \frac{\alpha}{2} (\tpsi d\phi - \phi d \tpsi) + \frac{\alpha}{2} (\psi d\tphi - \tphi d \tpsi)
\mid_{h}^2 + \\ \nonumber
 &+& 2 \pi~ 
\frac{e^{2 U} }{\rho^2}
\bigg( \mid D \phi \mid_h^2 + \mid D \psi \mid_h^2 
+ \mid D \tphi \mid_h^2 + \\ \nonumber
&+& \mid D \tpsi \mid_h^2  
+~\alpha ~( 
\mid D \psi \mid_h ~\mid D \tpsi \mid_h + \mid D \phi \mid_h ~\mid D \tphi \mid_h ) \bigg) \bigg],
\een
and by using the harmonic map connected with the extreme stationary axisymmetric solution in the considered theory $I(U',~\omega',~\phi', ~\tphi',~\psi',~\tpsi')$, one can show that \cite{chr09,cos10}
$$I(U',~\omega',~\phi', ~\tphi',~\psi',~\tpsi') \geq I(U,~\omega,~\phi, ~\tphi,~\psi,~\tpsi).$$
All these facts leads to the conclusion that
\be
m \geq \sqrt{
\frac{\mid J \mid^2}{m^2} + 2 \pi ~(Q^2_E + Q^2_M) + 2 \pi~ Q^2_{add}},
\label{m2}
\ee 
where by the quantity $Q^2_{add}$ we have denoted
\be 
Q^2_{add} = Q^2_{\tE} + Q^2_{\tM} + \alpha~ ( Q_E~Q_{\tE} + Q_M~Q_{\tM }).
\ee
The inequality can be rewritten in the form as
\be
m^2 \geq \frac{2 \pi~Q^2(total) + \sqrt{ 4 \pi^2~Q^4(total) + 4J^2}}{2},
\ee
where we defined $Q^2(total) = Q^2_E + Q^2_M + Q^2_{add}$. In terms of the above it can be concluded that the addition to the theory in question the {\it dark matter} sector, envisages the fact that
the mass with taking into account {\it dark matter} $(m_{dark~ matter})$ is greater than the mass in ordinary Einsten-Maxwell gravity
\be
m^2 _{dark ~matter} > m^2_{ordinary~ matter}.
\ee
The contemporary astronomical observations reveal that in our Universe there is approximately 7 times more {\it dark matter} than the ordinary one.
If we assume that the same situation holds for the {\it dark matter} charge, we can roughly estimate the mass of black hole in the theory with {\it dark matter} sector. It implies from the
equation (\ref{m2})
\be
m^2 _{dark ~matter}  \geq \frac{\mid J \mid^2}{m^2_{dark~ matter}}  + 2 \pi~(50 + 7\alpha)~Q^2_E + 2 \pi~(50 + 7\alpha)~Q^2_M.
\ee 
One can draw a conclusion that the aforementioned mass is significantly larger than for ordinary matter black object, almost fifty times greater.

In the Early Universe when the scaffolding of {\it dark matter} formed and visible matter condensated on it, it would be possible to emerge very massive black objects in a relatively short time after
Big Bang. This fact could explain the riddle connected with the possibility of growing such monsters. 

So far, roughly forty quasars with redshift greater than $z=6$ have been detected, each one of them contains a black hole with a mass of one billion solar masses.
The existence of such monsters is a tantalizing question for theory of black hole growth, formation, as well as, the coevolution of black objects and galaxies.
But there is only the tip of the black hole iceberg. Recently the detection of highly luminous
quasars associated with and powered by the accretion of material onto massive black hole envisages that there are \cite{fan13}-\cite{tho16}
black holes up to the ten billion solar mass, existed 13 billion years ago. The latest observations revealed such two fossils
of the dormant descendant of such population of black objects, at centers of Leo and Coma galaxy clusters.

\section{Area inequalities for dynamical black holes in Einstein-Maxwell dark matter gravity}
The next problem we pay attention to is the area inequalities for dynamical black holes in the theory in question. Consistently with our purpose, let us consider
a closed orientable two-dimensional spacelike surface $\cS$ smoothly embedded in the manifold under consideration.
The intrinsic geometry will be characterized by induced metric $q_{ab}$ with connection ${}^{(2)}D_a$,
Ricci scalar ${}^{(2)}R$, volume element $\ep_{ab}$, as well as, the area measure $d \cS$. 
Having in mind considerations presented in \cite{dai10,jar11,dai12,sim12}, one can introduce the outgoing and ingoing normal null vectors $l^a$ and $k^a$, respectively. They will be normalized to $-1$.
In the next step let us define the expansion $\theta^{(l)}$,
the shear $\si^{(l)}_{ij}$ and the normal fundamental form 
$\Omega^{(l)}_j$ which is connected with the outgoing null vector.
The aforementioned quantities are subject to the relations
\be
\theta^{(l)} = q^{ab}~\na_{a}l_b, \qquad
\si^{(l)}_{ij} = q^{c}{}{}_i~q^d{}{}_j~\na_{c} l_d - {1 \over 2}~\theta^{(l)}~q_{ij},\qquad
\Omega^{(l)}_j = - k^m~q^{r}{}{}_j~\na_m l_r.
\ee
One also assumes that the surface $\cS$ is stable and satisfies the marginally outer trapped surface condition, i.e., $\theta^{(l)} = 0$,
as well as we demand that the hypersurface in question is stable. It means that one can find an outgoing
vector $X_a = \la_1~l_a - \la_2~k_a$, with $\la_1 \ge 0$ and $\la_2 > 0$ satisfying
the condition of the form as 
$\delta_X \theta^{(l)} \ge 0$, where  the operator $\delta_X$ is the variation operator on surface $\cS$ along
the vector $X^a$ \cite{jar11}. 

As was mentioned in the preceding section, the surface ought to be axisymmetric,
due to the condition of the existence of the Killing vector field $\eta_a$.  On this account,  the following relations are also fulfilled:
\be
\cL_\eta l^j = \cL_\eta k^j = \cL_\eta \Omega^{(l)}_j = \cL_\eta F_{\alpha \beta} = \cL_\eta \tF_{\alpha \beta} =\cL_\eta B_{\alpha \beta} =\cL_\eta \tB_{\alpha \beta} =0.
\ee
It was revealed \cite{jar11} that for a closed marginally trapped surface $\cS$ satisfying the
stably outermost condition for vector $X^a$ and for every axisymmetric function $\talpha$,
one has that the following relation is provided:
\ben \label{in1}
\int_{\cS} d\cS ~&\bigg(& {}^{(2)}D_a \talpha~{}^{(2)}D^a \talpha + {1 \over 2}
\talpha^2~{}^{(2)}R \bigg) \ge
\int_{\cS} d\cS ~\bigg( \talpha^2~\Omega^{(\eta)}_j\Omega^{(\eta) j}
+ \talpha~\tbeta~\si^{(l)}_{ij}~\si^{(l) ij} \\ \nonumber
&+& G_{ab}~\talpha~l^a~(\talpha ~k^b + \tbeta~l^b) \bigg),
\een
where we have set $\tbeta = \talpha~\la_1/\la_2$.

The close inspection of the last term in equation (\ref{in1}) enables us to conclude that for Einstein-Maxwell {\it dark matter} gravity one arrives at
\ben
G_{ab}~\talpha~l^a~(\talpha ~k^b + \tbeta~l^b) 
&=& \frac{\talpha^2}{4} \bigg( \alpha E_m \tE^m + \alpha B_a \tB^a \bigg) + \frac{\talpha \tbeta}{2} ~\alpha~F_{im}~B_{ja}~q^{am}~l^il^j \\ \nonumber
&+& \frac{\talpha^2}{4} \bigg( E_m E^m + B_a B^a \bigg) + \frac{\talpha \tbeta}{2} ~F_{im}~F_{ja}~q^{am}~l^il^j \\ \nonumber
&+& \frac{\talpha^2}{4} \bigg( \tE_m \tE^m + \tB_a \tB^a \bigg) + \frac{\talpha \tbeta}{2} ~B_{im}~B_{ja}~q^{am}~l^il^j\\ \nonumber
&+& T_{ij}(matter)~\talpha l^i~(\talpha ~k^j + \tbeta~l^j).
\een
One ought to have in mind that the dominant energy condition for matter fields was assummed, i.e., $
T_{ij}(matter)\talpha l^i~(\talpha ~k^j + \tbeta~l^j) \ge 0$,
as well as null energy condition for
$U(1)$-gauge fields in the theory under inspection
\be
F_{ak}l^a~F_j{}^k l^j \ge 0,\qquad B_{ak}l^a~B_j{}^k l^j \ge 0, \qquad F_{ak}l^a~B_j{}^k l^j \ge 0.
\label{cond}
\ee
The conditions (\ref{cond}) lead to the conclusion that one receives positive terms on the right-hand side of the relation (\ref{in1}) (the second term in every line). Dropping these non-negative terms reveals
that the right-hand side of the inequality (\ref{in1}) can be rewritten in the form as follows:
\ben
\int_{\cS} d\cS ~\talpha^2~\bigg[
\Omega^{(\eta)}_j\Omega^{(\eta) j} &+& 
\frac{\alpha}{4}~\bigg( E_m \tE^m + B_a \tB^a \bigg) \\ \nonumber
&+& \frac{1}{4}~\bigg( E_m E^m + B_a B^a \bigg) 
+ \frac{1}{4}~\bigg( \tE_m \tE^m +  \tB_a \tB^a \bigg) 
\bigg].
\een
As in \cite{gab12}, in order to analyze the inequality in question, we introduce the axisymmetric line element on a two-dimensional hypersurface $\cS$
\be
ds^2 = q_{ab}~dx^a~dx^b = e^{\si}~ \bigg( e^{2 q}~d \theta^2 + \sin^2 \theta~d \varphi^2 \bigg),
\ee  
where $\si + q = const = c$. 
The fundamental form $\Omega^{(l)}_a$ can be decomposed using
the Hodge theorem in such a way that  
$\Omega^{(l)}_a = \ep_{ab}~D^b \omega + D_a \tla$.
On the other hand, the axisymmetry of $\Omega^{(l)}_a$ leads to the conclusions that 
$\Omega^{(\eta)}_a = 1 / 2 \eta~\ep_{ab}~D^b \omega$, where is connected with the gravitational part of the total angular momentum
$J = 1 / 8\pi \int_{\cS} dS~\Omega^{(l)}_a~\eta^a = (\omega(\pi) - \omega(0)) / 8$, while $\eta$ is given by $q_{ab}~\eta^a\eta^b$.

By the direct calculation it can be revealed that
$d\cS = e^c~dS_0$, where $ dS_0 = \sin \theta d\theta d \varphi$. 
Choosing $\talpha = e^{c - \si/2}$ we rewrite the inequality under inspection in the form as
\ben \label{m}
2(c &+& 1) \ge {1 \over 2 \pi}~\int_{\cS} dS_0~\bigg[
\si + {1 \over 4}~D_{m} \si D^m \si
 \\ \nonumber
&+& 
\frac{1}{  \eta^2}~\mid \bigg( 
D_j v + (\psi D_j \phi - \phi D_j \psi) + (\tpsi D_j \tphi - \tphi D_j \tpsi)  + \alpha (\tpsi D_j \phi - \phi D_j \tpsi)  + \\ \nonumber
&+& \alpha (\psi D_j \tphi - \tphi D_j \psi) 
\bigg) \mid^2 \\ \nonumber
&+& \frac{2 \pi}{ \eta}~\bigg(
D_a \psi D^a \psi + D_a \phi D^a \phi + D_a \tpsi D^a \tpsi + D_a \tphi D^a \tphi + \alpha (D_a \tpsi D^a \psi + D_a \phi D^a \tphi)
\bigg)
\bigg],
\een
where $\eta = q_{ij} \eta^i \eta^j$.
Since $A = 4\pi~e^c$, one can reach the inequality as follows:
\be
A \ge 4\pi~e^{\cM -2 \over 2},
\ee
where the functional $\cM$ is defined as the right-hand side of the equation (\ref{m}).

In order to prove the inequality among area, angular momentum and charges 
in Einstein-Maxwell {\it dark energy} theory,
one can consider the connection between
the functional $\cM$ and a harmonic energy for maps from the sphere into the complex hyperbolic space. 
The main point of the proof is to show that the extreme Kerr solution (with total charge $\tilde Q$) sphere, i.e., the set fulfilling
the Lagrange equations for the functional $\cM$ \cite{ace11}, realizes the absolute minimum of the functional $\cM$ in the set of all configurations having $J,~Q_E,~Q_{\tE}, ~Q_M,~Q_{\tM}$
fixed.

The theorem proposed in \cite{hil77} states that if the domain of any map is compact, bounded with non-void boundary as well as the target manifold has negative sectional curvature,
then a minimizer of the harmonic energy with Dirichlet boundary condition exists and moreover is unique, smooth and satisfies the Lagrange equations.  It follows that harmonic maps minimizers of the 
harmonic energy for given boundary Dirichlet conditions.

It can be proved that if $D = (\sigma,~\omega,~\psi,~\psi,~\chi,~\tchi)$ is a regular set on a sphere $S^2$
with fixed values of $J,~Q_E,~Q_{\tE},~Q_M,~Q_{\tM}$ then the following inequality is fulfilled:
\be
e^{\cM -2 \over 2} \geq 4 J^2 + {\tilde Q}^4(total),
\label {areamass}
\ee
where ${\tilde Q}^4(total) = (\sqrt{2\pi})^4~Q^4(total)$.\\
The proof of this statement involves a minimization problem and therefore 
can be attacked in different ways. One of them is to implement the connection between the integral $\cM$ and
a harmonic energy for mapping the sphere into the complex hyperbolic space. We use properties of the distance between harmonic maps in complex hyperbolic space. The
functional $\cM$ is related to the harmonic energy $\cM_D$
from a subset $D \subset S^2 \setminus \{\theta = 0,~\pi \}$ to the complex hyperbolic space, on which
the following line element is given
\ben
ds^2_H &=& \frac{d \eta^2}{\eta}
+ \frac{1}{\eta^2}~\bigg[ \bigg(
dv + (\psi d\phi - \phi d\psi) + (\tpsi d \tphi - \tphi d\tpsi)  + \alpha (\tpsi d\phi - \phi d\tpsi)  \\ \nonumber
&+& \alpha (\psi d_j \tphi - \tphi d_j \psi) 
\bigg) \mid^2  \bigg] \\ \nonumber
&+& \frac{2 \pi}{ \eta}~\bigg[
(d\psi)^2  + (d \phi)^2 + (d \tpsi)^2 + (d\tphi)^2 + \alpha (d \tpsi~ d\psi + d\phi~ d\tphi)
\bigg],
\een
while $\cM_D$ is provided by the expression
\be
\cM_D = \cM + \int_{\cS} dS~\ln \sin \theta + \int_{\p \cS}dl~(\si + \ln \sin \theta)~{\p \ln \sin \theta
\over \p n}.
\ee
We have set $n$ to be unit normal to the boundary to $\cS$ surface, while $dl$ is the measure
element of the boundary $\p \cS$. 

The proof of the statement is the same as in \cite{cle12}, therefore for the mathematical subtleties 
we refer the readers to this work. However for the readers' convenience we sketch the basic points of it.
Having in mind the theorem \cite{hil77}, we divide the sphere into three regions
\be 
\Omega_I = \{ \sin \theta \le e^{-(ln \ep)^2} \}, \qquad
\Omega_{II} = \{ e^{-(ln \ep)^2} \le \sin \theta \le \ep \}, \qquad 
\Omega_{III} = \{ \ep \le \sin \theta \}.
\ee
To commence with, one interpolates the potentials between extreme Kerr solution 
with {\it dark matter} total charge in $\Omega_I$ and
a general solution in $\Omega_{III}$ region. 
Let us remark that for the extreme Kerr black hole one has that
\be
m^2 = \frac{{\tilde Q}^2(total) + \sqrt{4 J^2 + {\tilde Q}^4(total)}}{2}.
\ee
If we concentrate on the non-extreme solution, the area of the bifurcation sphere can be provided by
\be
A(S_H) = 4 \pi~(r_H^2 + a^2) > 4 \pi \sqrt{4 J^2 + {\tilde Q}^4(total)}.
\ee
Because of the fact that mass tends to the mass of extremal stationary axisymmetric black hole solution, we get
\be
A(S_H)  \rightarrow 4 \pi \sqrt{4 J^2 + {\tilde Q}^4(total)}.
\ee
Hence, it leads to the Dirichlet problem in $\Omega_{IV} = \Omega_{II} \cup \Omega_{III}$, which
in turn implies that the mass functional of the extreme solution is less than or equal to the mass
functional for the additionally interpolating map on the whole sphere in question. Consequently, in the last step of the proof,  recalling that 
the limit as $\Omega_{III}$ converges to the sphere, 
and it shows that the considered mass functional for the additional map converges to the mass functional connected with the original sets. All the above
mathematics leads to the relation (\ref{areamass}).
One can remark that for two asymptotically flat ends, we can find an asymptotic stable ( for which the second variation of the area is nonnegative)  minimal surface
$\Sigma_{min} \in \cM$,  which separates two asymptotically
flat ends in question.  $\Sigma_{min}$ minimizes area among all the considered two-surfaces. It yields that 
$A(\Sigma_{min}) = A_{min}$, where $ A_{min}$ is the least area to enclose the ends \cite{dai13}. One obtains the relation as follows:
\be
A_{min} \ge 4 \pi ~\sqrt{4~J^2(\Sigma) + {\tilde Q}^4(total) (\Sigma)},
\label{area}
\ee
where in the above inequality $\Sigma$ stands either for $\Sigma_{min}$ or for $\Sigma_0$. The equality is satisfied when
$\Sigma = \Sigma_{0}$ and this case is responsible for the 
sphere of the extreme stationary axisymmetric solution in the theory in question.

\section{Inequalities for charge and size of the bodies}
In \cite{khu15} the inequalities relating charge and size of bodies in Einstein-Maxwell gravity were established. In general there are two attitudes to attack the problem in question.
Namely, one can assume the maximal case, i.e., $K_m{}^m$, or try to conduct the reasoning without this assumption. In this section we shall give some remarks concerning the size of the bodies
due to the existence of the {\it dark matter} sector. 

By the model of the body $\Theta$ we shall consider a connected open subset of the manifold with compact closure and smooth boundary.
From the equation of motion one can find that the square of the charge in Einstein-Maxwell {\it dark matter} gravity is provided by
\ben
Q^2 &=& \bigg( \frac{1}{4 \pi}\int_{\p \Theta} (E_\mu + \frac{\alpha}{2} \tE_\mu) n^\mu dS \bigg)^2 +
\bigg( \frac{1}{4 \pi}\int_{\p \Theta} (B_\mu + \frac{\alpha}{2} \tB_\mu) n^\mu dS \bigg)^2  \\ \nonumber
&+& \bigg( \frac{1}{4 \pi}\int_{\p \Theta} (\tE_\mu + \frac{\alpha}{2} E_\mu) n^\mu dS \bigg)^2 
+ \bigg( \frac{1}{4 \pi}\int_{\p \Theta} (\tB_\mu + \frac{\alpha}{2} B_\mu) n^\mu dS \bigg)^2 ,
\een
which reduces to the relation
\be
Q^2 \leq \frac{\mid \p \Theta \mid}{16 \pi^2} \int_{\p \Theta} dS~\gamma~\bigg(
E_a E^a + B_m B^m + \tE_a \tE^a + \tB_m \tB^m \bigg) + 2 \alpha~\bigg(
E_a \tE^a + B_m \tB^m \bigg),
\label{qq}
\ee
where we set
\be
\gamma = 1 + \frac{\alpha^2}{4}.
\ee
We add and extract the following term 
\be
\frac{\gamma~\mid \p \Theta \mid}{4 \pi^2}~\int_{\p \Theta} dS~\bigg( \mu - P_b \bigg),
\ee
to the right-hand side of the equation (\ref{qq}). Now by $\mu$ we have denoted
\be
\mu = \frac{1}{4} ~\bigg[
B_m B^m + E_a  E^a + \tB_m \tB^m +  \tE_a \tE^a
 + \alpha~(B_m \tB^m + E_a \tE^a) \bigg] + \mu_b,
\ee
Because of the fact that $\alpha $ is small, the relation for $Q^2$ implies
\ben
Q^2 &\leq& 
- \frac{\gamma~\mid \p \Theta \mid}{16 \pi^2}~\int_{\p \Theta} dS~\alpha~(1 - \frac{\alpha^2}{4} )^2~\bigg(
E_m \tE^m + B_c \tB^c \bigg)  \\ \nonumber
&-& \frac{\gamma~\mid \p \Theta \mid}{4 \pi^2}~\int_{\p \Theta} dS~\bigg( \mu_b - P_b\bigg) 
+ \frac{\gamma~\mid \p \Theta \mid}{4 \pi^2}~\int_{\p \Theta} dS~\bigg( \mu - P_b \bigg)  \\ \nonumber
&\leq& \frac{\gamma~\mid \p \Theta \mid}{4 \pi^2}~\int_{\p \Theta} dS~\bigg( \mu - P_b \bigg)  \leq \frac{\gamma~\mid \p \Theta \mid^2}{4 \pi^2}~ \mu,
\een
where we have used the dominant energy condition for the non-electromagnetic matter.

 Let us give some remarks concerning the size of the body. The crucial theorem was established in \cite{sch83}. To begin with one should define a closed curve in $\Theta$ which bounds a disk
 in this region of manifold in question. Let $q$ be the largest constant in the sense that the set of points within the distance $q$ of the aforementioned curve is contained in $\Theta$, as well as,
 forms a proper torus. Thus $q$ constitutes a measure of the size of the body with respect to the previously defined curve. The so-called Schoen-Yau radius $R_{SY}(\Theta)$ is connected with the largest
 value of $q$, one can find by examining all the curves (the Schoen-Yau radius can be expressed in terms of the largest torus embedded in $\Theta$). Having in mind these definition it can be proved that
 if we suppose that the body is a subset of the manifold with a scalar curvature ${}^{(3)}R$ bounded from below ${}^{(3)}R \geq \Lambda$ in $\Theta$, and we consider the maximal initial data, then
 \be
 \Lambda \leq \frac{8 \pi^2}{3}~\frac{1}{R^2_{SY}(\Theta)},
 \ee
 for a positive constant $\Lambda$.

Recalling he initial data equation (\ref{c2}) and the fact that $K_m{}^{m} = 0$, one can establish that ${}^{(3)}R \geq 16 \pi~\mu$, which in turn 
can be established as the value of a constant $\Lambda$.
Consequently, we arrive at the expression bounding $\mu$
\be
\mu \leq \frac{\pi}{6~R^2_{SY}(\Theta)}.
\ee
It leads to the conclusion that
\be
Q^2 \leq \gamma~\frac{\mid \p \Theta \mid^2}{2 \pi~R^2_{SY}(\Theta)}.
\ee
The modification of the above relations is to find the relations binding body parameters, without the assumption of vanishing the trace of the extrinsic curvature.
The reasoning is based on the reduction of the problem for general initial data to the maximal setting, i.e., in the maximal setting nonnegative scalar curvature is guaranteed from the 
point of view of the dominant energy condition.
Therefore one can deform the initial data metric to the new one, unphysical $\tg_{ab}$ for which ${}^{(3)}\tR >0$. 
In other words, we shall consider a graph $\tM = \{ t =f(x) \}$ inside a warped product of the manifold given by $M \times R;~\tg_{ab} = g_{ab} + dt^2$, with induced metric on $\tM$.

The manifold $\tM$ is referred to the so-called Jang surface and 
by the transformation of the metric tensor in the form as
$\tg_{ab} = g_{ab} + \na_a f \na_b f$, one achieves the desirable positivity of the scalar curvature.
The function $f$ satisfies the so-called Jang equation
\be
\bigg( g^{ab} - \frac{\na^af~\na^b f}{1 + \mid \na f \mid^2} \bigg) \bigg(\frac{\na_a\na_b f}{\sqrt{1 + \mid \na f \mid^2}} - K_{ab} \bigg) = 0.
\ee
The manifold $\tM$ is referred to the so-called Jang surface. 
However, the obtained scalar of the Jang graph is weakly positive, which means that in the defining relation there is a negative term \cite{khu15}.

One can measure the concentration of scalar curvature by estimating the first Dirichlet eigenvalue $\la_1$ of the operator ${}^{(\tg)} \na_\mu {}^{(\tg)} \na^\mu - 1/2 {}^{(3)}\tR $.
Let us assume that $\chi$ is the function related to the first eigenvalue $\la_1$. On this account, it is customary to write
\be
\la_1 = \frac{\int_{\Theta} d\Theta~\mid \na \chi \mid^2 + \frac{1}{2} {}^{(3)}\tR~\chi^2}{\int_{\Theta} d \Theta ~\chi^2}.
\ee
Having in mind that $\tR$ is weakly non-negative \cite{khu15}, we integrate by parts the above relation. 
Hence it leads to the relation
\be
\la_1 \geq 8 \pi~min_{\Theta}( \mu - \mid P \mid) = \Lambda. 
\ee
By virtue of the preposition \cite{sch83}, in which $\Lambda \neq 0$, we get
\be
\tR_{SY}(\Theta) \leq \sqrt{\frac{3}{2}} ~\frac{\pi}{\sqrt{\Lambda}},
\ee
where $\tR_{SY}$ is realated to the metric tensor $\tg_{ab}$. 
On the other hand, for an arbitrary positive function $\zeta \in C^{\infty}(\Theta)$, one obtains the following:
\be
\Lambda^{-1} \leq \frac{C_0}{8 \pi}~\frac{\int_\Theta d\Theta ~\zeta}{\int_\Theta~ d\Theta ~\zeta~(\mu - \mid P \mid)},
\ee
where
\be
C_0 = \frac{max_\Theta~(\mu - \mid P \mid)}{min_\Theta~(\mu - \mid P \mid)}.
\ee
The dominant energy condition on the boundary of the theory in question, can be given in one of the underlying forms, i.e.,
\be
\int_{\p \Theta} dS~(\mu_b - \mid P_b \mid) \geq 0, \qquad \int_{\p \Theta} dS~(\mu_b - \mid P \mid) \geq 0.
\ee
It enables to define a constant $C^2_1$ of the form \cite{khu15}
\be
C^2_1 = C_0 ~\frac{max_\Theta~(\mu - \mid P \mid)}{min_\Theta~(\mu - \mid P \mid)}.
\ee
Consequently we arrive at the following estimation of the charge of the body on the manifold in question:
\be
\mid Q \mid \leq \frac{1}{8}~C_1~\sqrt{\frac{3~\ga}{\pi}}~\mid \p \Theta \mid~\frac{1}{\tR_{SY}(\Theta)}.
\label{qbody}
\ee
As far as the formation of black hole due to the concentration of charge is concerned, the solution of the Jang equation has the irregularity, on the apparent horizon the function $f$ blows up.
The initial data have a boundary $\p M$, which will be considered as an outermost apparent horizon, i.e., each of the boundary components satisfies $\theta_+ = H_s + Tr~K =0$, for the future horizon
or $\theta_-= H_s - Tr~K =0$, for the past horizon. $H_s$ is the mean curvature with respect to the normal directed towards null infinity.

Thus, if we show that Jang equation has no regular solution, it can be stated that an apparent horizon is present in the considered initial data. It turned out that \cite{khu15}
the inequality (\ref{qbody}) can be rewritten as
\be
\mid Q \mid \geq \frac{1}{8}~C_1~\sqrt{\frac{3~\ga}{\pi}}~\mid \p \Theta \mid~\frac{1}{\tR_{SY}(\Theta)}.
\label{bhbody}
\ee
The above criteria to create black hole due to the concentration of  a charge of matter satisfying the dominant energy condition for the non-electromagnetic matter, reveals the fact that
the value of $\mid Q \mid $ is greater with respect to the influence of {\it dark matter}, i.e., the addition of {\it dark matter} sector modifies the relation for the charge of the body by a factor
which depends on the $\alpha$-coupling constant of the {\it dark matter}
$$\mid Q \mid (dark ~matter) = \sqrt{1 + \frac{\alpha^2}{4}} ~\mid Q \mid (ordinary~matter).$$

\section{Conclusions}
In our paper we have investigated the influence of the {\it dark matter} on angular momentum, mass and charge of the black holes. These parameters of the black hole
will be studied for axisymmetric maximal time-symmetric initial data for Einstein-Maxwell gravity with {\it dark matter} sector. {\it Dark matter} was simulated by another $U(1)$-gauge field
coupled with Maxwell one. The data set with two asymptotically flat regions, are given on smooth simply connected manifold, which enables to define potentials connected with fields in the theory,
as well as, to define charges for the adequate $U(1)$-gauge fields.

It turned out that the addition of the {\it dark matter} sector envisages the great increase of the mass of the black hole. Perhaps this fact may explain the observation of very massive black objects
in the Early Universe, when the visible matter condensate on the {\it dark matter} web.  We studied the area inequalities in Einstein-Maxwell {\it dark matter}
theory. The conclusion was that {\it dark matter} sector caused the increase of the area of the studied objects. Then, we proceed to discuss the connection between charge concentration and the size of the bodies.
We conduct our researches with the assumption of the maximal initial data and without it, using the so-called Jang equation.  Once more it turned out that the $\alpha$-coupling constant
binding the ordinary Maxwell field and {\it dark matter} gauge field, play the crucial role in the estimation of the charge-radius relations for body and black hole.


\begin{acknowledgments}
 MR was partially supported by the grant of the National Science Center $DEC-2014/15/B/ST2/00089$.\\
 \end{acknowledgments}


\end{document}